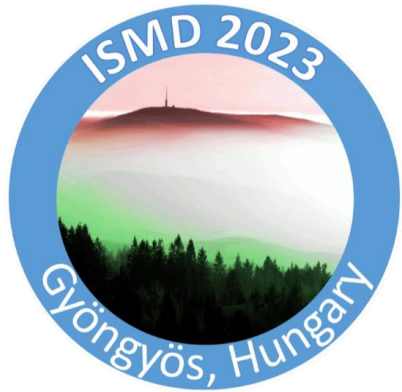

# A new game with Quark Matter Cards:

# The Eightfold path

Ana Uzelac[1]

[1]Zagreb, Croatia

December 27, 2023

**Abstract**

This paper presents a card game designed to educate and entertain players about the fundamental concepts of particle physics, particularly focusing on the classification of hadrons using the "Eightfold Path." The game, developed from the Quark Matter Card Games series, involves assembling elementary particle cards to form baryons and mesons on specific game boards. The rules and gameplay are detailed for two levels – beginner and intermediate. The game's educational value lies in reinforcing knowledge of particle properties, color neutrality, and symmetry principles that contributed to the development of the quark model and the Standard Model of particle physics. The physical background of the game is explored, highlighting the historical context of the "Eightfold Way" and its role in the emergence of the quark theory. The game is adaptable for different age groups and serves as a dynamic and engaging tool for learning complex physics concepts through entertainment. The paper concludes with acknowledgments to mentors and pioneers in card games with elementary particles.

# 1    Introduction

Csaba Török, a seventeen-year-old high school student in Hungary, on New Year's Eve 2008/2009, inspired by discussions about the physics of elementary particles, conceived the idea of creating a deck of cards depicting elementary particles, from which he developed the game "ANTI."

This game was later perfected and expanded, leading to the creation of a deck of cards with a set of games called „Quark Matter, Card Games" whose primary goal was entertainment, and secondary goals included the popularization of science and introducing players to basic concepts of particle physics. During the development of this game, Csaba teamed up with his peer Judit Csörgő. Tamás Csörgő, Judit's father and, as well as a research physicist engaged in experimental and theoretical high-energy physics related to the RHIC accelerator at Brookhaven, joined the team as a mentor and manager.

In the initial edition of *Quark Matter Card Games*, four different games are described that can be played using a deck of 66 cards representing some of the elementary particles (as well as antiparticles) of the Standard Model. [1]

In the deck of cards from [2], there are 66 cards, with each card representing a particle or antiparticle. Quarks and antiquarks are distinguished by colors, while leptons and antileptons are indicated by black-and-white cards.

The quarks represented on the cards are the up (*u*), down (*d*), and strange (*s*) quarks, and each of them can be colored in red, green, or blue, signifying the color charge of the individual quark. The leptons represented on the cards in the deck are leptons from the first and second generations, namely, the electron (e[-]), electronic neutrino ($\nu_e$), muon ($\mu^-$), and muonic neutrino ($\nu_\mu$). Additionally, the deck includes antileptons corresponding to these leptons, which are marked with plusses next to the particle names, as shown below[1]:

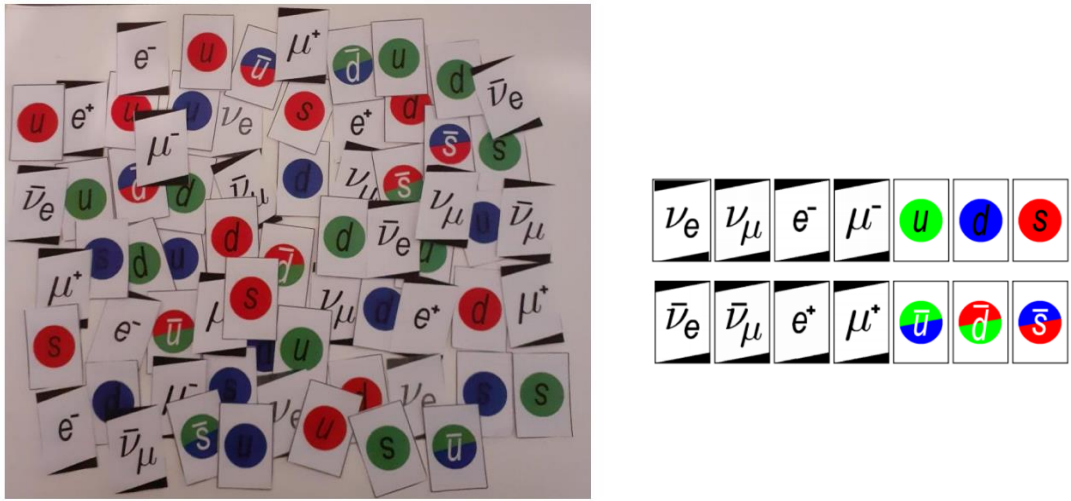

Figure 1 - Examples of different cards from the deck

---

[1] Pictures have been made using templates from [2]

Since then, several new games have been devised using the same deck of cards.

In these games, players with different levels of physics knowledge can not only have fun but also acquaint themselves with or enhance their understanding of various fundamental concepts in elementary particle physics. These include the names of elementary particles and antiparticles, the formation of baryons and mesons, the colors of quarks and hadrons, the electric charge of elementary particles, as well as baryon and lepton numbers and their conservation.[2]

Moreover, through such card games, players (individually or as part of a workshop, school, or science club) can delve into a wide range of additional topics. For instance, topics related to antiparticles and annihilation, the physics of the early universe, processes in accelerators and particle colliders, the so-called quark matter and quark-gluon plasma, particle decays and conservation laws, the phenomenon of cosmic rays, and the Higgs boson and experiments leading to its discovery.[2]. From the original deck of cards, entirely new and interesting ideas for games have emerged, such as Particle Poker[3] and the Rubik's Cube of Quark Matter. [1][4]

Due to the numerous interconnected themes in these games, they can serve as an excellent source for introducing (through entertainment) individuals to these seemingly complex and challenging areas of physics.

In my last year of studying physics education, inspired by all of these games that use a deck of cards with elementary particles, I decided to try my hand at designing new games with this deck, which could be related to aspects of elementary particle physics that were not already mentioned in existing games.

So I came up with the idea of a game which utilises the concept of *The eightfold path* for hadrons. This property of. "symmetry" in the representation of baryons and mesons was independently observed by Murray Gell-Mann and Yu'val Ne'eman in 1961.[6]

Their discovery played a crucial role in the development of elementary particle physics and nuclear physics. Namely, it was noticed that hadrons can be represented in suitable symmetric groups according to their charge (Q) and strangeness (S) properties. Due to the fact that such grouping often results in sets of eight (or more) hadrons, this property was named the "eightfold path" based on the Buddhist concept, and the formation of such a classification of hadrons also influenced the emergence and future development of the quark theory.

The game is intended for **two to four players** (due to the limited numbers of individual cards in the deck).

## 2   This is a (board) card game

For this game, in addition to the cards from the deck, the following boards are used (they can be drawn on a large piece of paper/cardboard or printed according to the following template)[5]:

**First board - "Meson Hexagon"**

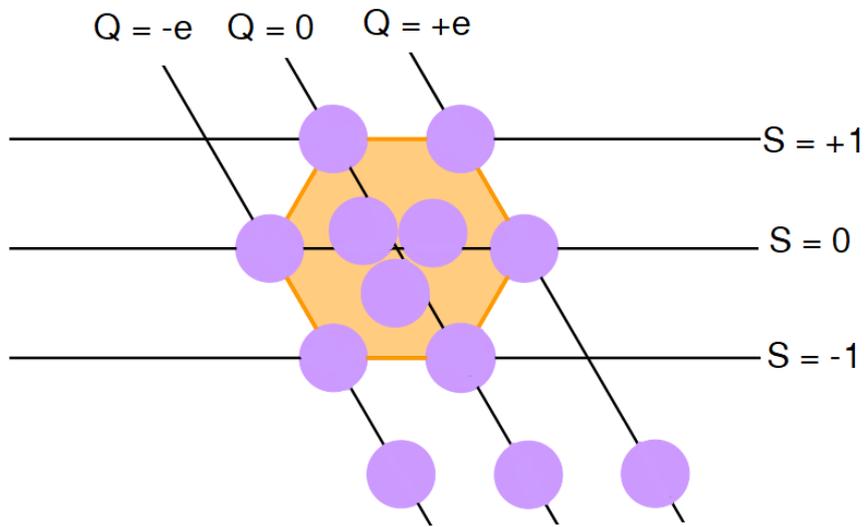

Figure 2 – Meson hexagon

Note: For the three mesons located in the center of the hexagon, their properties of total strangeness (S) and total electric charge (Q) are such as if they were at the very center of

the hexagon, where S=0 and Q=0. The symbol *e* represents the elementary charge; that is, the magnitude of the electric charge of an electron

**Second board - "Baryon Hexagon"**

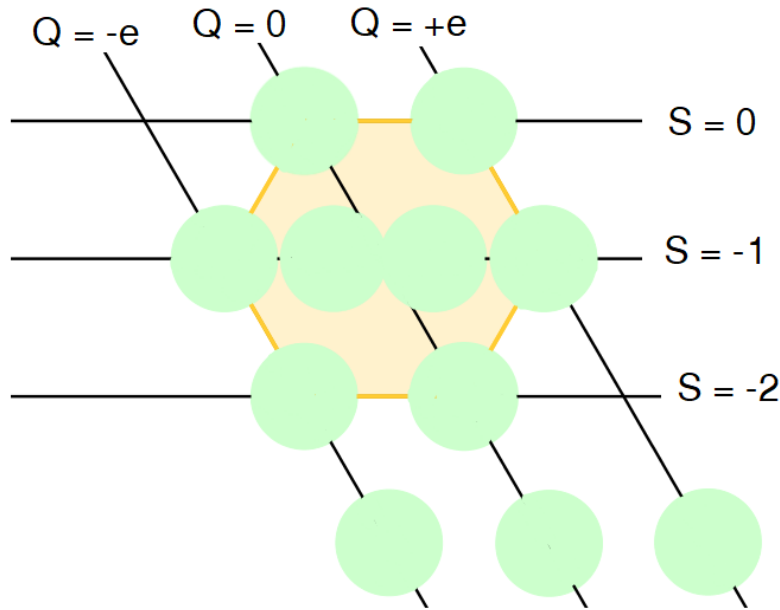

Figure 3 – Baryon hexagon

Note: For the two baryons located in the center of the hexagon, their properties of S and Q are such as if they were at the very center of the hexagon, where S=-1 and Q=0.

As can be observed, the boards correspond to the baryon and meson hexagons according to the concept of the "Eightfold Way" with three additional circles at the bottom.

The idea of the game is for players to fill the designated circles on the boards with the corresponding cards, following the rules of the game, so that within each circle, the total strangeness is equal to the one indicated with the capital letter S in the given row, and the total electric charge in each circle must be equal to the electric charge Q marked with diagonal lines. By doing that, players are arranging hadrons. Additionally, when arranging the corresponding hadrons, players must consider that they must be valid; meaning neutral in color.

The three lower circles on each picture, located outside the hexagon, represent circles that must be filled with 4 cards that can be leptons or antileptons (or their combination), so that their total electric charge is equal to the charge indicated on the corresponding diagonal lines.

There are two levels of this game possible:

- **Level 1 - Beginner:** In this level, names of baryons and mesons can be written on the boards. In that case, the game boards look like this[5]:

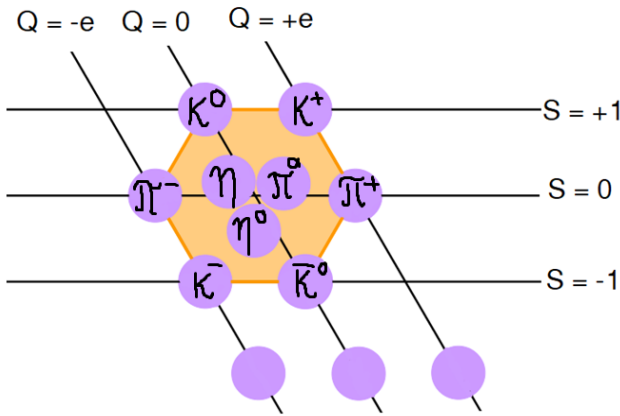
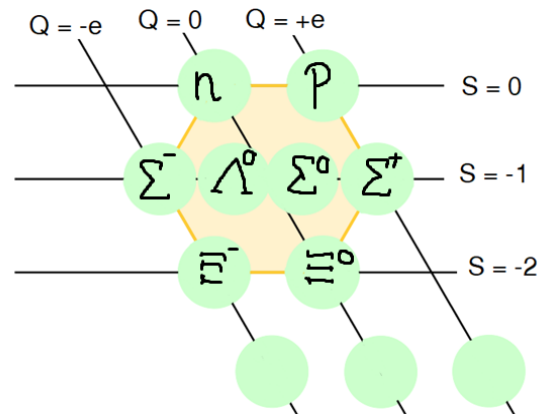

Figure 4 – "Mesonic Hexagon" adapted to a simpler version of the game.

Figure 5 – "Baryon Hexagon" adapted to a simpler version of the game

Also, during the game, players are allowed to use the tables provided below. Table 1 lists various baryons, and Table 2 lists different mesons. The tables display the names of particles and their symbols, along with the quark content of each. Symbols are used by physicists for a simplified notation of different particles, and it can be noted that the symbols of hadrons are marked with various letters from the Greek alphabet, with the upper index representing the charge of the hadron.[2]

| QUARK CONTENT | PARTICLE NAME | SYMBOL |
|---|---|---|
| *uuu* | Delta ++ | $\Delta^{++}$ |
| *uud* | Proton | $p^+$ |
|  | Delta + | $\Delta^+$ |
| *udd* | Neutron | $n^0$ |
|  | Delta 0 | $\Delta^0$ |
| *ddd* | Delta - | $\Delta^-$ |
| *uus* | Sigma + | $\Sigma^+$ |

| QUARK CONTENT | PARTICLE NAME | SYMBOL |
|---|---|---|
| uds | Lamda 0 | $\Lambda^0$ |
|  | Sigma 0 | $\Sigma^0$ |
| dds | Sigma - | $\Sigma^-$ |
| uss | Xi 0 | $\Xi^0$ |
| dss | Xi - | $\Xi^-$ |
| sss | Omega | $\Omega^-$ |

Table 1 – Baryons in Quark Matter Card Games[2]

| QUARK CONTENT | PARTICLE NAME | SYMBOL |
|---|---|---|
| $u\,\bar{d}$ | Positive pion | $\pi^+$ |
| $u\,\bar{u}$ | Neutral pion | $\pi^0$ |
| $d\,\bar{d}$ | Neutral pion | $\pi^0$ |
| $\bar{u}\,d$ | Negative pion | $\pi^-$ |
| $u\,\bar{s}$ | Positive kaon | $K^+$ |
| $d\,\bar{s}$ | Neutral kaon | $K^0$ |
| $\bar{d}\,s$ | Neutral anti-kaon | $\bar{K}^0$ |
| $\bar{u}\,s$ | Negative kaon | $K^-$ |
| $s\,\bar{s}$ | Eta meson | $\eta^0$ |

Table 2 – Mesons in Quark Matter Card Games[3]

---

[2] The table has been created according to [2]
[3] The table has been created according to [2]

- **Level 2 - Intermediate:** In this level, players do not have pre-written hadrons on the board that they need to assemble; instead, they must place them on the board themselves based on their total electric charge Q and strangeness S.

For this level, players need to be familiar with the individual strangeness and charge of each elementary particle in the deck, as the total strangeness S and electric charge Q of each hadron (or group of leptons) is equal to the sum of strangeness (s) and electric charge (q) of all individual particles that make it up.

The values for each type of card in the deck are listed in the following table[5]:

| *Particle name* | Strangeness (S) | Electrical charge (Q / e) |
|---|---|---|
| $u$ | 0 | +2/3 |
| $d$ | 0 | -1/3 |
| $s$ | -1 | -1/3 |
| $\underline{u}$ | 0 | -2/3 |
| $\underline{d}$ | 0 | +1/3 |
| $\underline{s}$ | +1 | +1/3 |
| $e^-, \mu^-$ | 0 | -1 |
| $e^+, \mu^+$ | 0 | +1 |
| $\nu_e, \underline{\nu}_e, \nu_\mu, \underline{\nu}_\mu$ | 0 | 0 |

Table 3 – Values of electric charge and strangeness for cards in the deck

In this version of the game, players need to quickly and successfully handle the addition of the specified values for individual cards to obtain the total strangeness and electric charge. They must also be skilled at recognizing potential "opportunities" to assemble the appropriate hadron that is left unfilled on the board.

The actual gameplay and all rules are explained below.

## 3 Rules and course of the game

At the very beginning of the game, it is necessary to divide the cards from the deck into three groups – the first group should contain leptons and antileptons, the second group antiquarks, and the third group quarks. After dividing the cards into the appropriate groups, it is essential to thoroughly shuffle the cards in each of them, and then place the cards from each group on the table with their faces down. This way, a row of three groups of cards is formed on the table:

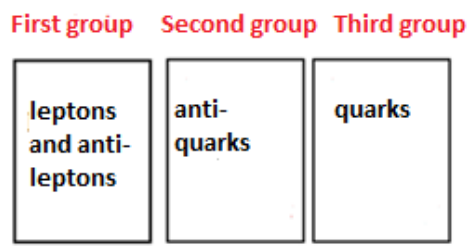

Figure 6 – Initial Card Formation. The cards are arranged according to the type of elementary particles, but they are faced down, so the players cannot see them.

At the beginning of the game, each player must take two cards from each pile. Accordingly, each player must have six cards at the beginning of the game - two (anti)leptons, two antiquarks, and two quarks. Players hold their cards in their hands so that only they can see them. Then, the first part of the game begins, using the "Mesonic Hexagon" board!

**FIRST PART OF THE GAME - ASSEMBLING ANTILEPTONS, LEPTONS, AND MESONS**

The game is played **in turns**, so players need to agree on the order in which they take their turns – for example, clockwise. In each turn, each player has the right to choose **one of the following two options**:

a) In this option, the player can choose one card (whichever they prefer) from their hand and return it to the BOTTOM of the corresponding pile (for example, a lepton can only be returned to the pile designated for leptons and antileptons). Then, the player can draw one new card from the TOP of any other pile (of course, the player doesn't know exactly which card they will draw, but they can choose whether they want to draw a (anti)lepton, a quark, or an antiquark).
After such card exchange, the player loses their turn, and the next player takes their turn.

b) If the player has the appropriate cards in their hand, they can place them in the corresponding unfilled spots on the board. Within each circle, the cards must be placed in one move - either a meson (composed of a quark and an appropriate anti-quark of the respective colors) is placed in one of the suitable positions inside the hexagon, or four cards that are a combination of leptons and anti-leptons with a charge equal to the one required in the corresponding circle outside the hexagon.

After the player places the cards on the board, they must take as many cards from the central piles as the number of cards they placed on the board, so that at the end of the turn, they have six cards in their hand again. **They can choose how many cards to take from each pile!**

In this case, the rule applies that the cards are taken from the top of each group, so the player cannot know exactly which cards they will draw, but they can choose the type of particles they want to take.

Each placement of a corresponding meson or a group of four corresponding (anti)leptons on the board earns the player 1 point. If a player places a group of cards that is not appropriate - in other words, it does not represent a valid meson or a group of leptons whose properties match those specified in a particular circle, the player who points out the mistake earns one point, and the player who made the mistake receives one negative point, after which the incorrectly placed cards are returned to the bottom of the corresponding piles.

This turn-based game continues until the entire board is filled with appropriate cards.

An example of the final situation for the first part of the game is shown in the following figure:

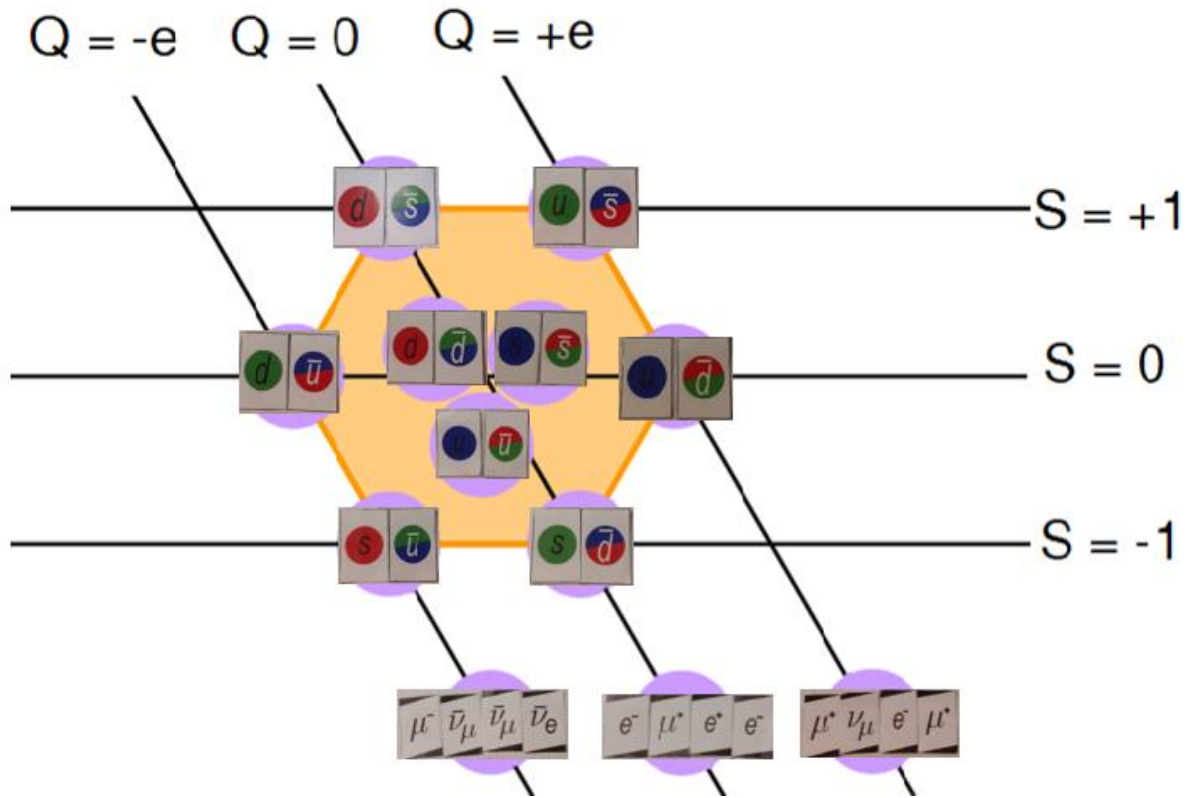

Figure 7 – Example of a correctly filled board at the end of the first part of the game

As can be seen, the central circles contain mesons that are composed of pairs of a quark and its own antiquark, and for each meson on the edges, it is true that its own anti-meson is located on the edge opposite to it.

After completing the first part of the game, the second part of the game follows, in which the "Baryon Hexagon" board is used.

**SECOND PART OF THE GAME - ASSEMBLING ANTILEPTONS, LEPTONS, AND BARYONS**

After the first part of the game is finished, players must return all their remaining cards, as well as all the cards that were on the "Meson Hexagon" board, to their respective piles. After that, the cards in each pile need to be thoroughly shuffled and placed on the table, just like in the initial part of the previous game.

However, in this part of the game, the pile with antiquarks is not used, so the table initially contains the following **two groups of cards** (face down)[5]:

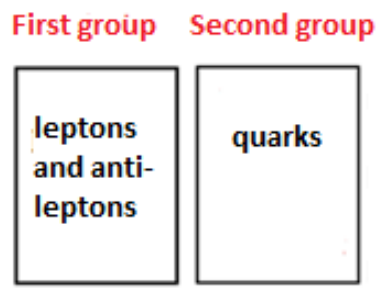

Figure 8 - Groups of cards for second part of the game

In this part of the game, players take turns one by one, and during each turn, a player has the right to one of the following **two options** (which are very similar to the rules in the previous game):

a) The player can return one card (whichever they choose) from their pile back to the BOTTOM of the corresponding pile and, instead, take one new card from the TOP of any other pile, just like in the previous game with assembling mesons.

After such a card exchange, the player loses their turn, and the next player takes their turn.

b) Just like in the previous part of the game, if a player has the appropriate cards in their hand, they can place them on the appropriate vacant spot on the board. Cards within each circle must be placed in one move - meaning that either an appropriate baryon (composed of three quarks of different colors) or four cards that are a combination of leptons and antileptons with the required charge must be placed on the board at once.

Similarly, after the player places cards on the board, they must take as many cards from the central piles as they placed on the board so that they end their turn with the same number of cards as before. The player can choose how many cards they will take from each pile!

In this case, each completion of a circle on the board, i.e., placing the appropriate baryon or a group of four (anti)leptons in the correct spot on the board, earns the player 1 point. Also, just like in the previous game, each mistake made by a player results in one negative point for that player and one point for the person who notices the mistake.

In this way, the game continues until the entire board is filled with the corresponding cards, for example, as shown below:

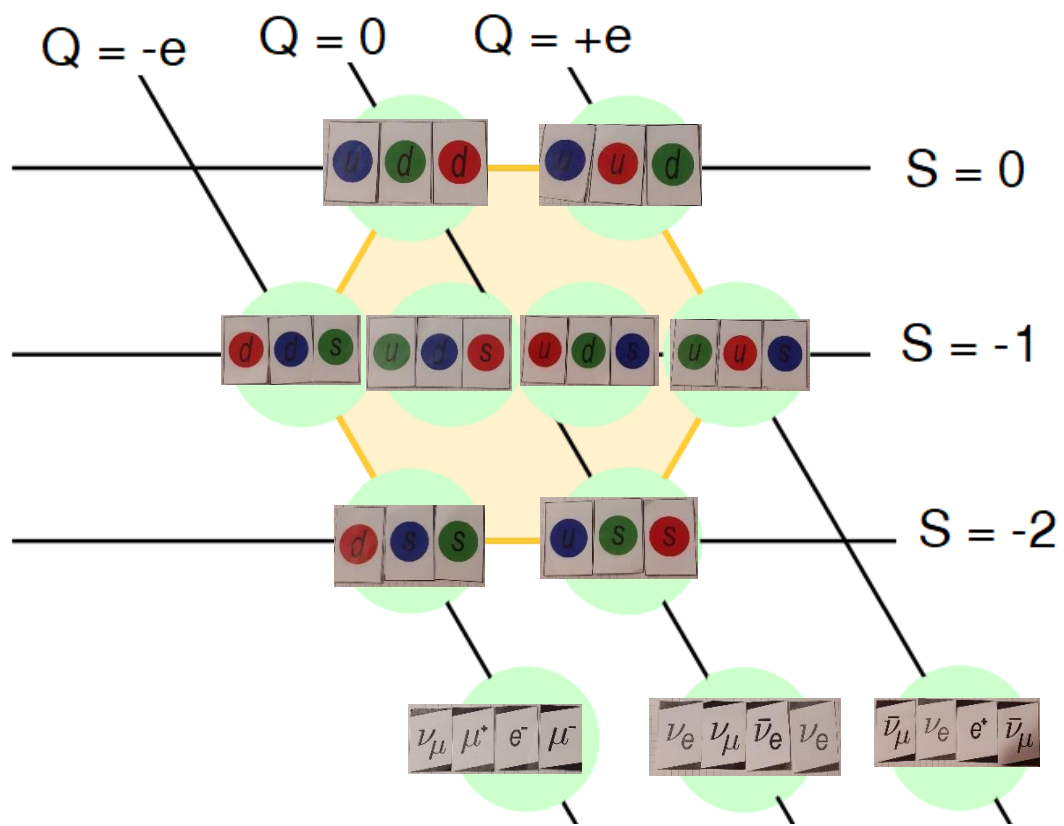

Figure 9 – Example of a correctly filled board at the end of the second part of the game

After all the circles on this board are filled with the corresponding cards, the game is over, and the player who has accumulated the **highest total number of points** from both parts of the game is the winner!

## 4  For whom is the game intended?

The simpler level of the game is suitable for all players who play and understand other similar games in which hadrons are formed from quarks (for example, players familiar with the intermediate level game "Quark Matter").

While playing, players can reinforce their knowledge of these hadrons and their color neutrality, but they can also discover something new – that they can be organized into symmetric groups (in this case, hexagons) based on their total electric charge and strangeness properties, which is actually the fundamental idea of this game from which the quark model of hadrons was developed, along with many properties of particle symmetries that form the foundation of the Standard Model. By understanding the fundamental concepts of this game and its influence on the further development of physics, an awareness of physical reasoning and the historical development of science is fostered.[5]

After players have grasped all the rules and mastered the game at the simpler level, they can move on to the more complex level, which requires knowledge of the properties of elementary particles listed in tables they could use. In this intermediate level of the game, players need to skillfully handle several different "variables"; they need to constantly consider the total sum of individual electric charges and strangeness of each card, and organize and manage their cards and drawing new ones to successfully fill the required empty spaces.

Because of this, the game may not be suitable for very young children (under 12 years old) who do not yet have developed abstract thinking and logical reasoning, or basic knowledge of these quantities.

However, older children and adults without extensive prior knowledge of these topics can still master this game and find it just as enjoyable as those who know more about particle physics. The outcome of the game depends on an element of luck since players do not know which cards they will exactly draw from different sets, but it also depends on quick thinking, concentrated and active engagement in the game, and knowledge of particle properties. Players can gain an advantage by properly exchanging a card that is not needed in their deck with a card from another set that is required to successfully form certain hadrons

Thus, the game is dynamic and interesting, the results are unpredictable, but a good understanding of physics, along with actively tracking possible combinations and skillfully using the right opportunities, are highly desirable!

As mentioned earlier, this game is based on the physical concept of the "Eightfold Path" in representing hadrons. It involves recognizing symmetrical patterns, such as those used on the game boards, to depict groups of different baryons and mesons. The development of this idea was very important in the further development of quark theory and the Standard Model, and more about it can be read in the chapter „Physical background of the game - "Eightfold path" in the representation of hadrons", which is included in another QR code in this poster.

 If we were to present this game to students as part of a workshop, then after mastering the basic techniques, a group discussion prompted by questions based on the given excerpt would be desirable.

# 5 Physical background of the game - "Eightfold path" in the representation of hadrons

The discovery of neutral kaons in 1947, positively charged kaons in 1949, and lambda particles in 1950 revealed a peculiar property of these particles. Namely, all of them could be easily produced in particle collisions (in accelerators and, for example, during cosmic ray showers), but after their creation, it was observed that they decayed much more slowly than expected.

For instance, in a study of cosmic rays in 1947, it was found that the lambda particle, which was created in the collision of two protons, had a lifetime of approximately $10^{-10}$ seconds, instead of the predicted $10^{-23}$ seconds (based on the mass of this particle and its strong interaction).[7]

In 1953, Murray Gell-Mann, Tadao Nakano, and Kazuhiko Nishijima independently proposed the postulation of a new quantum number, which was named "strangeness" (S), and which was conserved during the creation but not during the decay of such hadrons, thus explaining their "peculiar" property.[9]

During the 1950s and 1960s, with the development of the first particle colliders and detector improvements, numerous new particles were discovered. The number of newly discovered particles became so large that they began to be categorized into three different groups, depending on their mass and mode of interaction. The lightest particles (like electrons) were called leptons, moderately heavy particles (like pions) were called mesons, and the heaviest ones (like protons and neutrons) were called baryons. Initially, all these particles were thought to have no internal structure and were considered "elementary" particles.

In 1961, Gell-Mann and Yu'val Ne'eman independently noticed that numerous hadrons could be organized into symmetric groups based on their strangeness quantum number (S) and charge (Q). [8] It involved a system of organizing different hadrons into regular geometric patterns, which, due to the fact that they often form groups of eight hadrons, was collectively named the "Eightfold Way."

For example, the eight lightest baryons can be represented in a hexagonal pattern as shown in the following figure[5]:

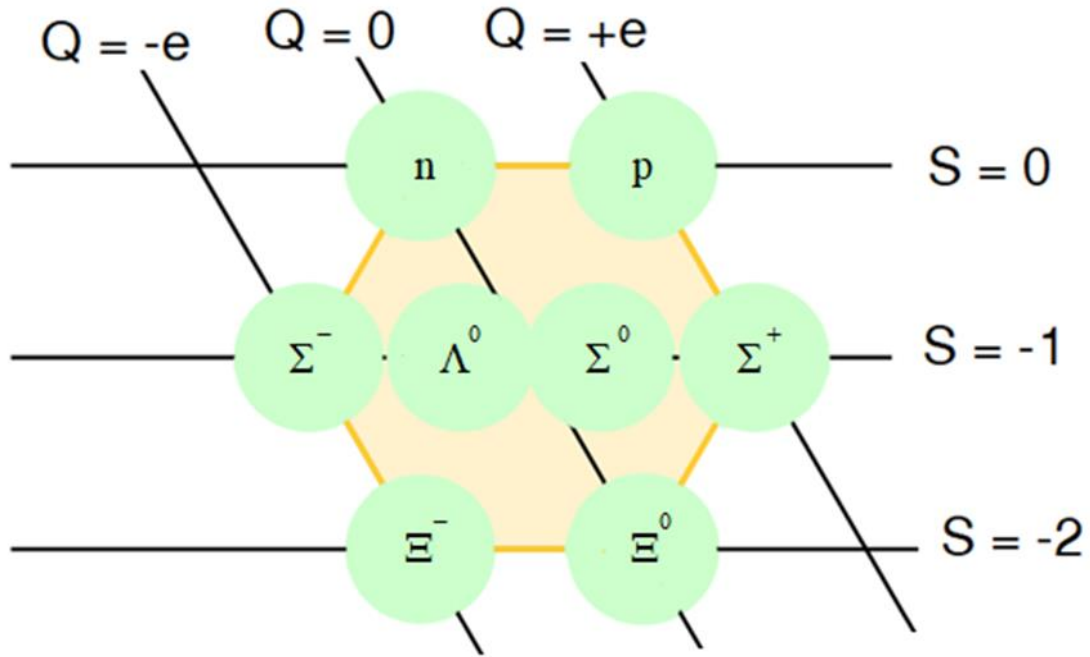

Figure 10 – Example of a symmetric classification of baryons

As is evident, the horizontal lines represent lines of constant "strangeness" quantum number (S), and on the slanted lines, there are hadrons with equal charge (Q).

Moreover, each of the mentioned eight baryons has a baryon number of 1, spin ½, and positive parity.

From this mode of organization, it was observed that there exists a relationship between the masses of individual particles.[10]

$$m(n) - m(\Xi^0) = \frac{1}{2}[3m(\Lambda^0) + m(\Sigma^0)] \qquad \text{Equation 1}$$

$$m(\Xi^-) - m(\Xi^0) = m(\Sigma^-) - m(\Sigma^+) + m(p) - m(n) \qquad \text{Equation 2}$$

Considering that numerous other hadrons could be represented in similar octets, it was discovered that their masses also obey the aforementioned equations if they are included in a way that the mass of each particle in other octets is placed in the equation corresponding to its position in the octet shown in Figure 9.[5]

For example, in an entirely identical manner, the following nine mesons can be represented, where their strangeness is +1, 0, or -1, electric charge -e, 0, or +e, and their spin is all equal to zero:

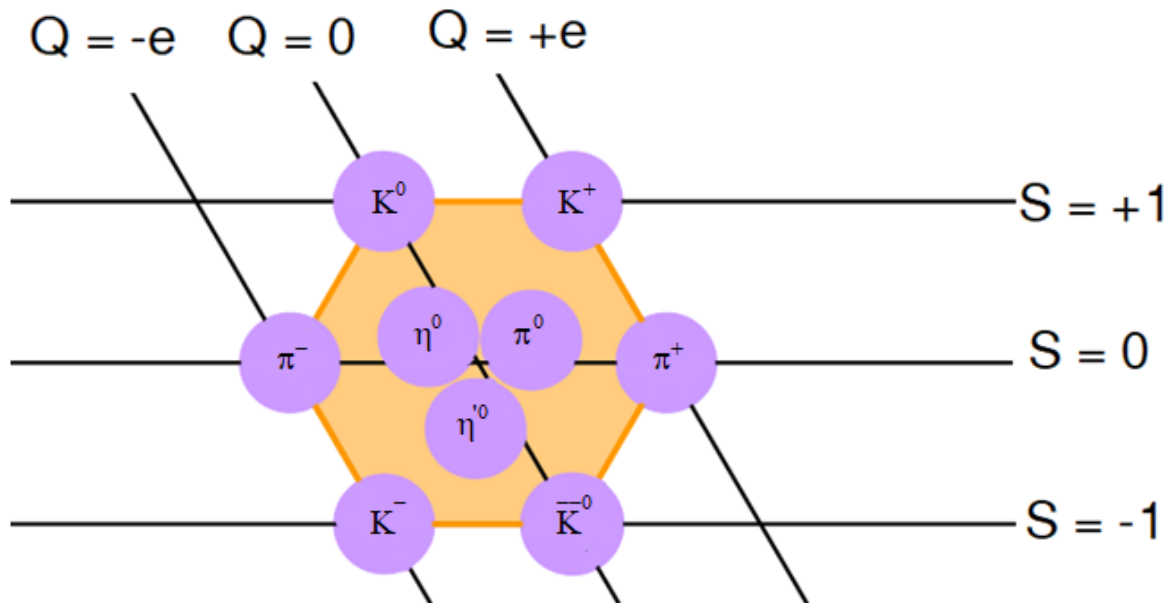

Figure 11 – Example of a symmetric classification of mesons

Similarly, based on their electric charge and strangeness values, the heavier baryons with a spin of 3/2 can be organized into symmetric "triangular" patterns, as shown in the following example

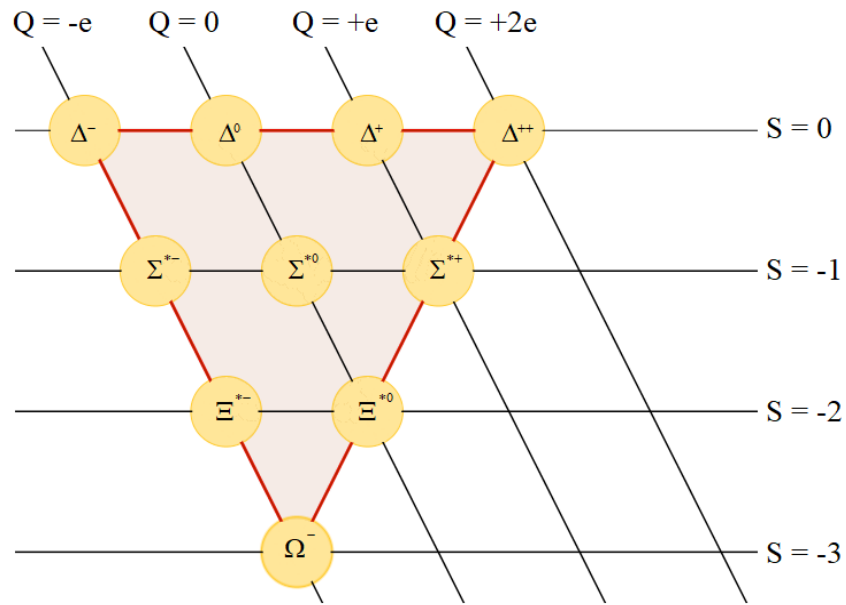

Figure 12 – Example of a symmetric classification of heavier baryons

In 1961, when this method of organization was observed, the particle at the bottom of the "triangle" (denoted by Ω⁻) had not yet been discovered! However, Gell-Mann predicted its existence based on these symmetric patterns and inferred its properties, such as its rest mass (1672 MeV), spin ($\frac{3}{2}$), strangeness (-3), and electric charge (-*e*).[10]

Based on these predicted data, appropriate experiments were designed to search for such a particle, and in 1964, the particle called Ω⁻ was indeed detected,[8] confirming the validity of this classification approach!

This method of predicting unknown particles and their properties is reminiscent of Mendeleev's approach, where, based on empty spaces in the periodic table of elements, previously unknown chemical elements and their properties could be predicted.

Another important insight from the concept of the "Eightfold Way" is the origin of the strong force. Upon examining this model, it was noted that just as the electric force between two charged bodies is transmitted by exchange particles called photons, the strong force must be transmitted between certain other smaller charged particles through the exchange of eight possible particles.[10]

Following such insights, the beginnings of the theory of quarks were developed in 1964, based on the idea of Murray Gell-Mann and his colleagues. According to this theory, quarks, which are particles with a spin of $\frac{1}{2}$ and interact through the strong force, are the fundamental building blocks of mesons and baryons.

If we closely examine the quark content of the hadrons depicted in the last three figures, we can observe that in all these cases, they are hadrons constructed from a combination of three quarks - u, s, or d. It is precisely by considering these hadrons and their symmetry properties that the initial quark theory emerged in 1964, predicting the existence of these three quarks!

Therefore, it is evident that the idea of the "Eightfold Way" significantly influenced the emergence of the Standard Model and the development of particle physics as we know it today.

# 6 Conclusion

Through the described card games, various ideas and concepts of particle physics, which may seem very complicated at first, can be brought closer to players of different ages and backgrounds through fun and social interaction. Therefore, these games show great potential in popularizing particle physics.

In addition to existing games, there is a particularly interesting and significant possibility of creating new and diverse games centered around the existing deck of cards, which could introduce and teach additional topics. As a result, these games can continue to be developed in parallel with new physical discoveries and theories, thereby reaching different groups of people.

# 7 Acknowledgments

Thanks and credits are due to prof dr. sc. Dubravko Klabučar, a professor at Theoretical Physics Division of Particles and Fields at PMF Faculty of Science, Department of Physics : As a mentor for my master's thesis, he introduced me to the quark matter card games and provided exceptional support and assistance throughout the thesis preparation, and later on. Additionaly, I would like to express my gratitude to prof. dr. Tamás Csörgő from MATE

University in Hungary for the support and encouragement; together with prof. Klabučar, he encouraged me to me write this article, and to participate in the ISMD conference.

In conclusion, special thanks to all the aforementioned pioneers of card games with elementary particles, who have been a tremendous source of inspiration. Without their creativity and efforts, my ideas would not even exist.